\documentclass[11pt]{article}
\usepackage{amssymb}
\usepackage[preprint]{acl}

\usepackage{times}
\usepackage{latexsym}
\usepackage{booktabs}
\usepackage{makecell}

\usepackage{environ}
\usepackage[most]{tcolorbox}
\usepackage{enumitem}
\usepackage{lipsum}
\usepackage[table]{xcolor} 

\usepackage[T1]{fontenc}
\usepackage[utf8]{inputenc}

\usepackage{microtype}

\usepackage{inconsolata}

\usepackage{graphicx}
\usepackage{amsmath}

\title{The Sonar Moment: Benchmarking Audio-Language Models \\ in Audio Geo-Localization}

\author{
 \textbf{Ruixing Zhang\textsuperscript{1}},
 \textbf{Zihan Liu\textsuperscript{1,2}},
 \textbf{Leilei Sun\textsuperscript{1,3}},
 \textbf{Tongyu Zhu\textsuperscript{1,3}},
 \textbf{Weifeng Lv\textsuperscript{1}}
\\
 \textsuperscript{1}The State Key Laboratory of Complex and Critical Software Environment, Beihang University \\
 \textsuperscript{2}Shanghai AI Laboratory, \\
 \textsuperscript{3}The Key Laboratory of Data Science and Intelligent Computing, International Innovation Institute, Beihang University
\\
}

\begin{document}

\maketitle

\begin{abstract}

\begin{figure*}[t]
\begin{center}
\centerline{\includegraphics[width=2\columnwidth]{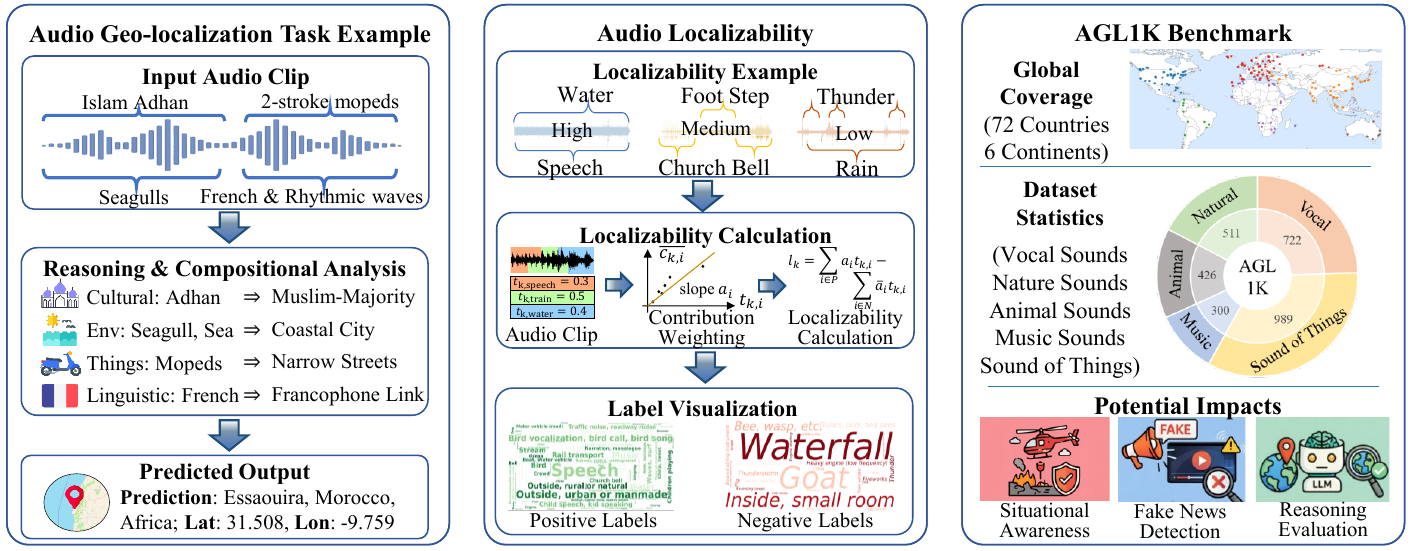}}
\caption{Overview of AGL1K, illustrating the audio geo-localization task, localizability, and composition.}
\label{figure1}
\end{center}
\end{figure*}

Geo-localization aims to infer the geographic origin of a given signal. In computer vision, geo-localization has served as a demanding benchmark for compositional reasoning and is relevant to public safety. In contrast, progress on audio geo-localization has been constrained by the lack of high-quality audio-location pairs. To address this gap, we introduce AGL1K, the first audio geo-localization benchmark for audio language models (ALMs), spanning 72 countries and territories. To extract reliably localizable samples from a crowd-sourced platform, we propose the Audio Localizability metric that quantifies the informativeness of each recording, yielding 1,444 curated audio clips. Evaluations on 16 ALMs show that ALMs have emerged with audio geo-localization capability. We find that closed-source models substantially outperform open-source models, and that linguistic clues often dominate as a scaffold for prediction. We further analyze ALMs’ reasoning traces, regional bias, error causes, and the interpretability of the localizability metric. Overall, AGL1K establishes a benchmark for audio geo-localization and may advance ALMs with better geospatial reasoning capability.

% short版本如下

% Geo-localization seeks to infer the geographic origin of a signal. While prior work has largely focused on images, progress on audio geo-localization has been limited by the lack of large-scale, high-quality audio-location pairs. We introduce AGL1K, the first world-scale dataset for audio geo-localization, covering 72 countries and territories. To identify reliably localizable samples from crowd-sourced recordings, we propose an Audio Localizability metric that quantifies geographic informativeness, resulting in 1,444 curated audio clips. Evaluations on 16 audio language models show that the task is tractable yet challenging. Closed-source models substantially outperform open models, and linguistic clues often dominate prediction. We further analyze model reasoning behavior, regional bias, error distributions, and the interpretability of the localizability metric. AGL1K establishes a benchmark for studying audio geo-localization and advancing ALMs with stronger compositional reasoning and world knowledge.

\end{abstract}

\vspace{0.5em}
\begin{flushleft}
\small
\textbf{Code:} \url{https://github.com/Rising0321/AGL1K} \\
\textbf{Space:} \url{https://huggingface.co/spaces/RisingZhang/AudioGeoLoc}
\end{flushleft}

\section{Introduction}

Geo-localization aims to infer the geographic origin of a signal and offers a compelling alternative to standard classification because it requires compositional reasoning over diverse clues. Mapping observations to a single GPS coordinate not only requires perceptual inference but also broad world knowledge of geography and culture. In computer vision, this problem is typically studied as image geo-localization~\cite{classification1,retrival1,vlm1}. As for audio, it can also provide analogous evidence, including reverberation patterns, traffic density, and coastal wave dynamics. Beyond academic interest, audio geo-localization has clear societal value. For example, assisting investigations of audios disseminated by extremist actors can help fact-checkers verify the claimed location of viral content. These applications make audio geo-localization relevant to public safety. However, despite this potential, \textbf{a systematic benchmark for audio geo-localization remains absent}.

The lack of audio geo-localization benchmarks stems from two main factors. First, \textbf{there is no publicly available audio dataset with location annotations}. In contrast to image geo-localization, where progress has been enabled by large-scale geo-tagged data from social media platforms, no comparable resource exists for audio. Second, \textbf{the field lacks a quantitative notion of audio localizability to filter for geographically informative recordings}. Even if crowd-sourced platforms can provide large numbers of audio-location pairs, without such a measure it remains nontrivial to identify samples that carry meaningful geographic signals.

To systematically evaluate audio geo-localization capability in audio language models (ALMs), we introduce AGL1K, the first audio geo-localization benchmark for ALMs. AGL1K is curated from the crowd-sourced Aporee platform and filtered using our proposed Audio Localizability metric, which estimates the geographic informativeness of each recording by aggregating evidence from both positive and negative sound categories during inference. The resulting benchmark spans 72 countries across six continents and covers diverse acoustic scenes, including nature soundscapes, animal vocalizations, music, human-made sounds, and spoken conversations. \textbf{This diversity makes AGL1K a suitable testbed for assessing compositional reasoning in modern ALMs, with potential downstream relevance to public safety and misinformation detection.}

Our comprehensive evaluation of 16 ALMs shows that current ALMs have begun to exhibit meaningful audio geo-localization capability, yet AGL1K remains challenging. The results reveal a clear capability hierarchy between closed- and open-source models. Leading ALMs (e.g., Gemini 3 Pro) demonstrate strong knowledge and reasoning, with failures increasingly driven not by missing information but by over-commitment to a single clue. In contrast, open-source models exhibit more fundamental limitations in fine-grained perception, which prevents reliable extraction of geographically informative signals.

Through detailed error analysis, we conclude three findings suggesting directions for improving future audio models. (1) \textbf{Enhance fine-grained perception}: Many open-source ALMs misidentify languages and other subtle acoustic clues, limiting perceptual sensitivity and downstream knowledge retrieval. (2) \textbf{Mitigate regional bias:} Systematic prediction imbalance such as over-predicting certain continents or regions, persists across models. (3) \textbf{Strengthen compositional reasoning}: As audio geo-localization requires integrating multiple weak clues, models must avoid relying on any single clue as decisive evidence.

In summary, this work makes the following contributions:

\begin{itemize}

\item \textbf{The first audio geo-localization benchmark for ALMs.} We introduce AGL1K, a benchmark comprising 1,444 user-uploaded audio clips from 72 countries, covering diverse acoustic scenes including nature soundscapes, animal vocalizations, music, human-made sounds, and spoken conversations.

\item \textbf{A principled notion of Audio Localizability.} We propose an Audio Localizability, a quantitative measure of an audio's geographic informativeness. The metric aggregates informativeness from positive and negative sound labels during inference, enabling filtering localizable recordings.

\item \textbf{A comprehensive evaluation of state-of-the-art ALMs.} We benchmark 16 ALMs and find the emergence of audio geo-localization to some extent. We further analyze representative models' reasoning traces, regional biases, and error causes to provide future insights for improving ALMs.

\end{itemize}

% 在应用上，本文提出的方法未来可以对在互联网上发布极端视频的嫌疑人进行定位与抓捕，并对虚假信息进行核验，会对公共安全也产生重要的意义。

% Beyond benchmarking, this work introduces a new lens for analyzing world knowledge in ALMs and provides insights into their emerging geospatial reasoning capabilities, with implications for future research on multimodal cognition.

\section{Related Work}

\subsection{Geo-Localization}

% Geo-localization aims to infer geographic locations from real-world signals and is fundamental to applications such as security, navigation, and environmental monitoring. 
Most existing geo-localization research focuses on image-based geo-localization, which can be broadly categorized into three paradigms. 
\textbf{Classification-based} approaches discretize the Earth's surface into predefined regions, but achieving fine-grained accuracy requires a large number of classes, making training and scalability challenging~\cite{classification1,classification2,classification3,classification4}. 
\textbf{Retrieval-based} methods embed images and geographic coordinates into a shared representation space and localize by nearest-neighbor search. However, they rely on massive, globally distributed, annotated image databases that are costly to construct and maintain~\cite{retrival1,retrival2,retrival3,retrival4}. 
More recently, \textbf{vision-language model}-based methods exploit the world knowledge encoded in large multimodal models to directly predict locations, demonstrating promising performance~\cite{vlm1,vlm2,vlm3}.
In contrast, audio geo-localization remains unexplored. Existing benchmarks are limited in diversity, often focusing on narrow domains such as bird vocalizations~\cite{audiogeo}.

\subsection{Audio-Language Models}

% Audio Language Models(ALMs)的研究随着深度学习的发展而发展。在深度学习兴起之时，Deep Speech 2模型用端到端神经网络替代了复杂的自动语音识别流水线，在英文和中文上取得接近人工转录水平的性能。紧接着，大规模自监督语音表示学习取得了重要进展，如Whisper在零样本条件下对多语言的语音转录达到与有监督方法相当的水准。随着大模型技术的发展，语音已经成为了first-class modality within unified architectures in pretraining。许多大模型natively support audio understanding and generation within unified architectures including GPT4o~\cite{gpt4o} and Qwen3-Omni~\cite{qwen3o}.

% Audio-Language Models (ALMs) have emerged as a core paradigm for multimodal reasoning. Existing approaches can be broadly categorized into three stages.
% \textbf{Representation alignment} methods freeze both the audio encoder and large language model, training only a lightweight alignment module, as exemplified by Audio Flamingo~\cite{Audio-Flamingo}. 
% \textbf{Embedding-level audio tokenization} incorporates audio at the representation level that share the embedding space with text, such as Qwen-Audio~\cite{Qwen-Audio}. 
% More recently, \textbf{Omni models} integrate audio as a first-class modality within unified architectures in pretraining, natively support audio understanding and generation within unified architectures, rather than fusing audio into a pretrained text-only LLM. Notable examples include GPT4o~\cite{gpt4o} and Qwen3-Omni~\cite{qwen3o}.

The development of ALMs has closely followed the advances in deep learning. 
Early breakthroughs such as Deep Speech 2~\cite{deepspeech} replaced traditional multi-stage pipelines with end-to-end neural networks, achieving near-human transcription performance. 
Subsequent progress in self-supervised and weak-supervised learning further advanced the field, exemplified by Whisper~\cite{whisper}, which demonstrated multilingual transcription performance comparable to supervised methods. 
With the rise of large model pretraining, audio has emerged as a first-class modality within unified architectures. Recent foundation models such as GPT-4o~\cite{gpt4o}, Gemini3~\cite{gemini3}, Qwen3-Omni~\cite{qwen3o}.

A wide range of benchmarks has been proposed to evaluate ALMs, covering basic Speech-to-Text Translation (OpenAudioBench~\cite{OpenAudioBench}), Vocal Sound Classification (VocalSound~\cite{vocal}), as well as higher-level reasoning tasks such as spatial reasoning (Star-Bench~\cite{star}) and causal discovery (MECAT~\cite{causal}). 
However, there is still no dedicated benchmark for assessing audio geo-localization ability in ALMs.
% However, despite this progress, there is still no dedicated benchmark for assessing compositional geographic reasoning in the audio domain.

\section{Audio Geo-Localization Benchmark}

% 现在思路应该是我们要做composition的reasoning，我们该如何得到这么一个benchmark。raw 数据 -> 数据处理 -> 最后得到这个数据

The Audio Geo-Localization Benchmark (AGL1K) is designed to evaluate the geographic compositional reasoning capabilities of audio-language models. We next describe how we obtain large-scale audio-location pairs (\ref{accus}), perform coarse quality control (\ref{filter}), identify reliably localizable recordings using our proposed Audio Localizability metric (\ref{loc}) and post-processing after obtaining ALMs' response (\ref{post}).

\begin{figure*}[t]
\begin{center}
\centerline{\includegraphics[width=2\columnwidth]{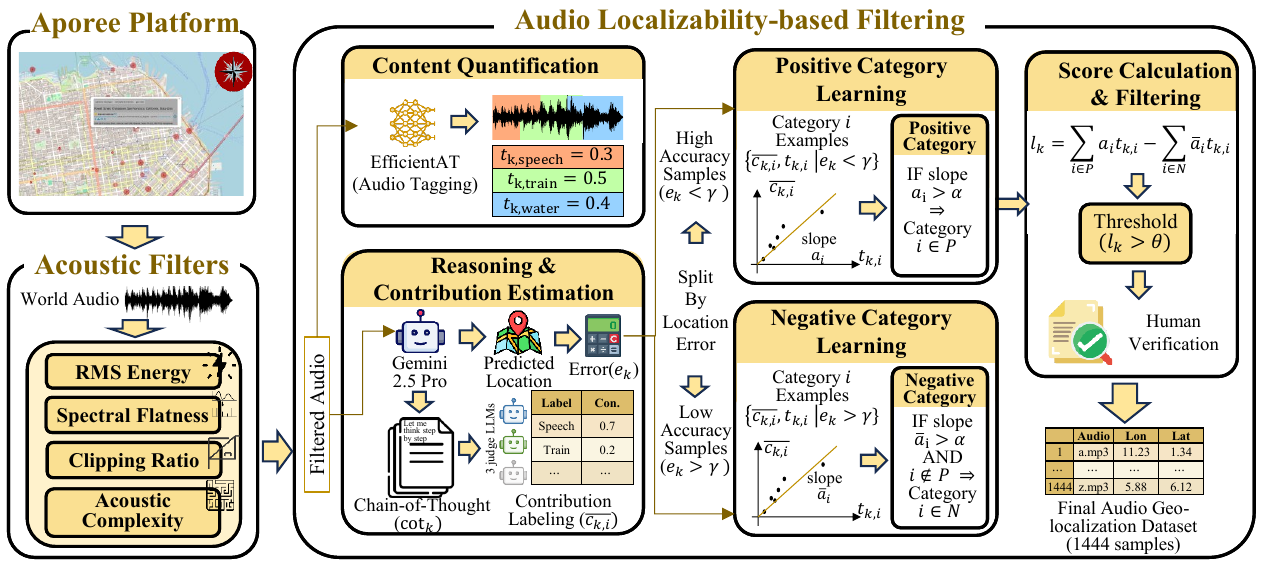}}
\caption{\textbf{Overview of the benchmark construction framework.} AGL1K is curated from the crowd-sourced Aporee platform. The recordings are first filtered using four acoustic filters, followed by our proposed Audio Localizability measure, which quantifies the geo-informativeness of each sample.}
\label{figure3}
\end{center}
\end{figure*}

\subsection{Dataset Acquisition}
\label{accus}
Global-scale audio geo-localization has received limited attention, largely due to the scarcity of GPS-tagged audio recordings. We identify Aporee\footnote{\url{https://aporee.org/maps/}}, a crowd-sourced platform where users upload geo-tagged audio worldwide. In collaboration with the Aporee team, we obtain tens of thousands of audio-location pairs released under an Apache license. Additional details about Aporee are provided in Appendix~\ref{apore}.

\subsection{Initial Filtering} 
\label{filter}
Because Aporee recordings are user-uploaded, their quality varies substantially. To coarsely remove low-quality or uninformative samples, we apply four acoustic filters targeting common failure causes: low signal energy, noise-like content, clipping, and overly monotonous recordings. Specifically, we compute (1) \emph{RMS Energy} to exclude extremely low-amplitude audio, (2) \emph{Spectral Flatness} to remove noise-like signals, (3) \emph{Clipping Ratio} to discard heavily clipped recordings, and (4) \emph{Acoustic Complexity} to eliminate clips with limited temporal variation. These filters retain recordings with sufficient acoustic structure and potential geographic clues. Formal definitions are provided in Appendix~\ref{app:fils}.

\subsection{Audio Localizability}
\label{loc}

\subsubsection{Localizability Calculation}
While the initial acoustic filters remove low-quality recordings, they do not ensure that the remaining samples contain enough geo-informativeness evidence. To construct a benchmark with localizable inputs, we introduce a principled measure of \textbf{audio localizability}.

We model the localizability of an audio recording as the net contribution of \emph{informative} (positive) versus \emph{uninformative} (negative) sound categories. Intuitively, longer exposure to informative categories (e.g., language, place-specific human activities) should improve geo-localizability, whereas prolonged uninformative content (e.g., infant crying) can obscure relevant clues. Categories also contribute unequally: coast sounds may provide coarse evidence, while linguistic content often offers finer specificity. Motivated by this, we define the localizability of sample $k$ as
\begin{equation}
l_k = \sum_{i \in P} a_i t_{k,i} - \sum_{i \in N} \bar{a}_i t_{k,i}.
\label{eq1}
\end{equation}
where $P$ and $N$ denote the sets of positive and negative sound categories, respectively; $t_{k,i} \in [0,1]$ is the fraction of time that category $i$ is present in sample $k$; and $a_i$ (resp.\ $\bar{a}_i$) quantifies the contribution strength of category $i$ when it acts as a positive (resp.\ negative) category. We obtain $t_{k,i}$ by tagging each recording with EfficientAT~\cite{efficientat} under the AudioSet ontology~\cite{audioset}. Estimating category-level contributions is therefore central to assessing localizability.

Because audio geo-localization lacks established expert heuristics, we derive category contributions from the reasoning behavior of a strong ALM. For each sample $k$, we run a Gemini 2.5 to produce a predicted location and a chain-of-thought $cot_k$, and compute the distance error $e_k$ between the prediction and the ground-truth coordinates. We then prompt three language models to assess, based on $cot_k$, the contribution of each detected audio category to the model's reasoning, using a five-level discrete scale. Averaging across the three judges yields $\bar{c}_{k,i}$, the estimated contribution of category $i$ for sample $k$.

To determine which category acts as a positive versus a negative category, we fit simple linear relationships between category duration and judged contribution. For samples with $\{k|e_k < \gamma \}$, we fit $y = a_i x$ with $x = t_{k,i}$ and $y = \bar{c}_{k,i}$, and include category $i$ in $P$ if $a_i > \alpha$. For samples with $\{k|e_k \geq \gamma \}$, we fit $y = \bar{a}_i x$ and include category $i$ in $N$ if $\bar{a}_i > \alpha$ and $i \notin P$. Using Equation~\ref{eq1}, we deem a recording highly localizable if $l_k > \theta$.

Applying the above criterion yields a pool of highly localizable recordings. From this pool, we manually curate 1,444 high-quality clips, balanced between samples with and without human speech, to form the final dataset used in our audio geo-localization benchmark. We set $\alpha = 1/3$ ,$\gamma = 1000$~km and $\theta = 1$ empirically.

\subsubsection{Top Positive and Negative Categories}
\begin{figure}[h]
\begin{center}
\centerline{\includegraphics[width=\columnwidth]{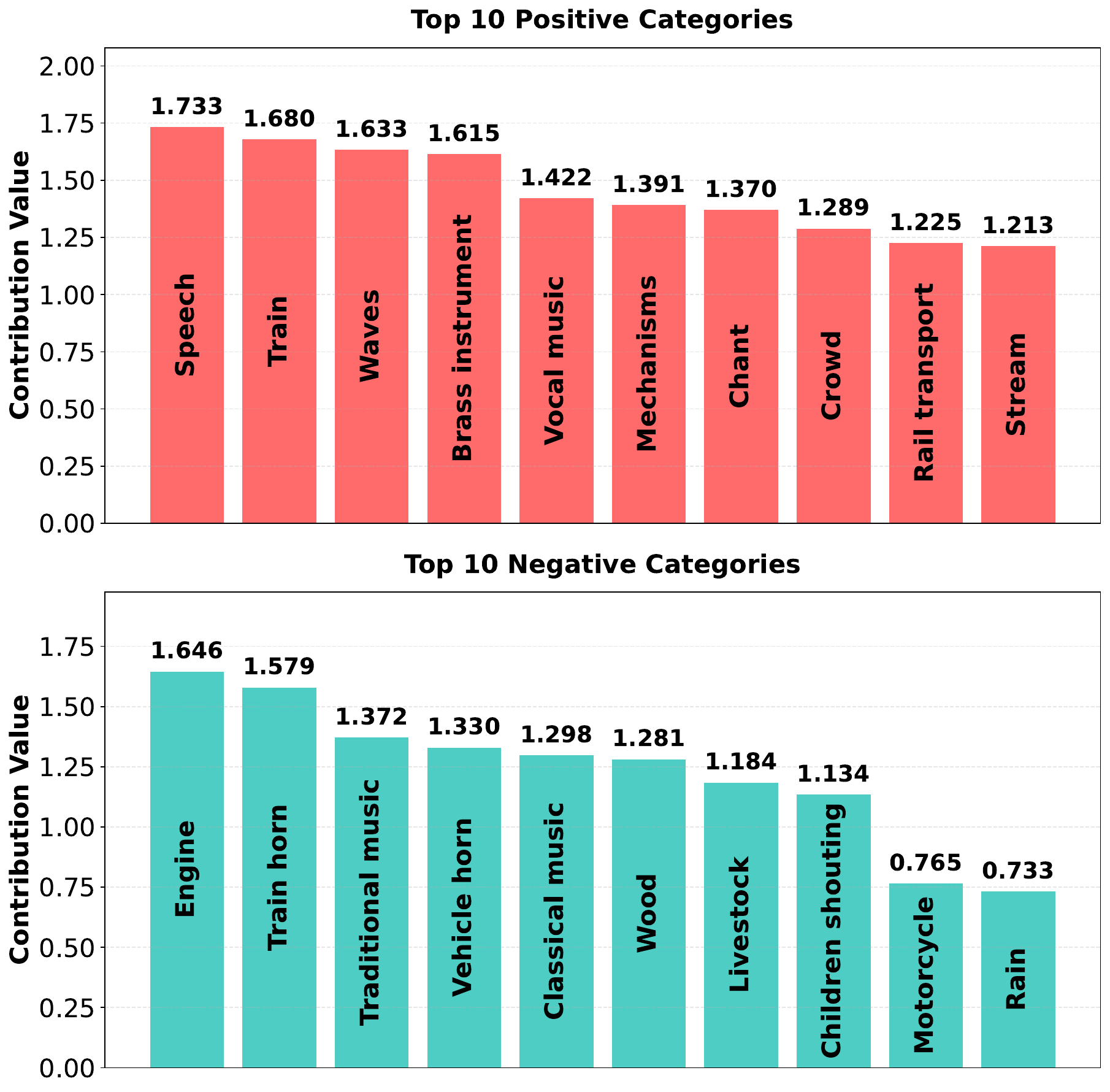}}
\caption{\textbf{Top Positive and Negative Categories.}}
\label{fig_cloud}
\end{center}
\end{figure}

% To evaluate the quality of the proposed localizability metric, we present the top 10 positive and negative sound labels (Figure~\ref{fig_cloud}). Several clear patterns emerge: \textbf{Speech} is the dominant positive feature, reflecting the high geographic specificity of linguistic clues. 而对于rail trransport，waves这种明显具有地理分布的数据，也具有较高的localizability。然而，对于Engine，Train horn这样明显的噪音，或是Rain，Wood这样在世界上分布极广的声音，其就具有较高的反向contribution。

To evaluate the quality of the proposed localizability metric, we present the top 10 positively and negatively contributing sound categories (Figure~\ref{fig_cloud}). 
Several clear patterns emerge: Speech is the most prominent positive contributor, highlighting the strong geographic specificity encoded in linguistic cues. 
Additionally, sounds with inherently regional distributions such as rail transport and waves also exhibit high localizability.
In contrast, generic noise sources like engine or train horn, as well as globally pervasive natural sounds such as rain and wood, tend to have strong negative contributions, likely due to their ubiquity and limited geographic discriminability.
Therefore, \textbf{these results demonstrate that the learned feature attributions align well with human intuition, supporting the interpretability and validity of the localizability metric.}

% To evaluate the quality of the proposed localizability metric, we present the top 10 positive and negative sound labels (Figure~\ref{fig_cloud}). 
% Several clear patterns emerge: \textbf{Speech} is the dominant positive feature, reflecting the high geographic specificity of linguistic clues. 
% Outdoor environments, both urban and rural, are consistently associated with positive contributions, while indoor scenes correlate negatively. 
% In addition, sound categories with explicit geographic relevance, such as traffic, motor vehicles, and ocean sounds, are assigned positive weights, whereas acoustically distinctive but geographically uninformative entities, such as pigs or fireworks, receive negative weights.
% \textbf{These results demonstrate that the learned feature attributions align well with human intuition, supporting the interpretability and validity of the localizability metric.}

\subsubsection{Localizability Examples}
\begin{figure}[h]
\begin{center}
\centerline{\includegraphics[width=\columnwidth]{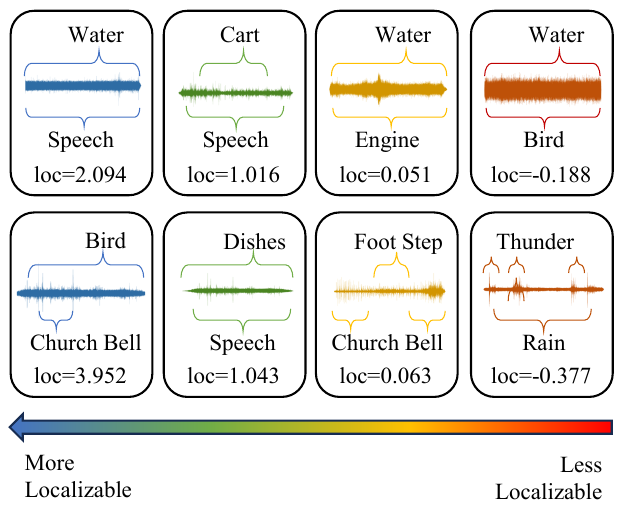}}
\caption{\textbf{Localizability Examples.} Localizability increases from right to left.}
\label{fig_loc}
\end{center}
\end{figure}

% Figure \ref{fig_loc} 在这里展示了8个样本计算出的localizability。可以看到仅有thunder和rain的声音具有最低的localizability，可能因为这会发生在世界上绝大多数国家。而尽管footstep和church bell有较低的可定位性，如果听到了鸟叫，其中的可定位性会大幅增加，因为鸟的种类可能能提供较高的声音线索。而对于第二列的声音片段，尽管存在speech，但由于充斥着cart，disher等室内声音，使得他们的可定位性有较大的削弱。综上，, the localization aligns with human inference knowledge，providing strong backbone to filter the data.

Figure~\ref{fig_loc} visualizes the computed localizability for eight example clips. Recordings dominated by thunder and rain yield the lowest scores, likely because these sounds are ubiquitous across most regions worldwide. Although clips containing footsteps or church bells are also weakly localizable, the presence of birdsong can substantially increase localizability, as species-specific vocalizations provide stronger geographic clues. For the clips in the second column, speech is present, but it is heavily masked by indoor foreground noises (e.g., carts and dishware), which markedly weakens geographic informativeness. Overall, \textbf{the localizability scores are consistent with human intuition and support our metric as a reliable basis for filtering localizable data.}

% Audio Geo-Localization performance of different models.  closed-source models和open-source models里最好的结果分别加粗了，次好的结果分别加下划线了.
\begin{table*}[t]
  \centering
  \small
  \setlength{\tabcolsep}{3pt} % 调小列间距，适配两栏
  \caption{Audio geo-localization performance of different models. The best results within the closed-source and open-source model groups are separately bolded, while the second-best results are underlined in each group.}
  \label{tab:audio_geoloc}
  \begin{tabular}{lrrrrrrrrrrrr}
    \toprule
    Model &
\makecell{Geo-\\score$\uparrow$} &
\makecell{Distance$\downarrow$} &
\makecell{Cont.\\acc.$\uparrow$} &
\makecell{Country\\acc.$\uparrow$} &
\makecell{City\\acc.$\uparrow$} &
\makecell{Reject\\rate$\downarrow$} &
\makecell{$<\!1$\\acc.$\uparrow$} &
\makecell{$<\!10$\\acc.$\uparrow$} &
\makecell{$<\!500$\\acc.$\uparrow$} &
\makecell{Speech\\dis.$\downarrow$} &
\makecell{Non-speech\\dis.$\downarrow$} \\
    \midrule
    RANDOM                 & 112.16 & 9869.01 & 0.14   & 0.01   & 0.00   & 0.00   & 0.00 & 0.00 & 0.00 & 9930.88 & 9717.65 \\
    \midrule
    \multicolumn{12}{l}{\textbf{Closed-Source Models}} \\
    \midrule
    Gemini 3 Pro           & \textbf{3031.82} & \textbf{2180.57} & \textbf{0.82} & \textbf{0.51} & \textbf{0.11} & \underline{0.01} & \textbf{0.07} & \textbf{0.19} & \textbf{0.52} & \textbf{1548.29} & \textbf{3727.33} \\
    Gemini 2.5 Pro         & \underline{2826.95} & \underline{2521.97} & \underline{0.78} & \underline{0.49} & \textbf{0.11} & \textbf{0.00} & \underline{0.05} & \underline{0.17} & \underline{0.48} & \underline{1634.22} & 4693.67 \\
    Gemini 2.5 Flash       & 2256.22 & 3558.37 & 0.65 & 0.39 & 0.07 & \underline{0.01} & 0.02 & 0.12 & 0.36  & 2689.96 & 5682.74 \\
    Gemini 2.5 Flash-Lite  & 1687.97 & 4373.89 & 0.55 & 0.28 & 0.03 & 0.02 & 0.01 & 0.06 & 0.21 & 3742.97 & 5917.32 \\
    Gemini 2.0 Flash-Thinking & 2494.46 & 2991.51 & 0.73 & 0.39 & 0.07 & \underline{0.01} & 0.02 & 0.12 & 0.38  & 2265.47 & 4767.61 \\
    Gemini 2.0 Flash       & 2535.49 & 2906.31 & 0.73 & 0.40 & 0.08 & \underline{0.01} & 0.02 & 0.13 & 0.41  & 2189.30 & \underline{4660.31} \\
    Gemini 2.0 Flash-Lite  & 2417.14 & 3223.85 & 0.71 & 0.38 & 0.06 & 0.04 & 0.01 & 0.11 & 0.37 & 2296.28 & 5492.96 \\
    GPT-4o Audio Preview   & 1996.15 & 4067.87 & 0.61 & 0.37 & 0.05 & 0.04 & 0.02 & 0.10 & 0.30 & 2841.03 & 7069.09 \\
 
    \midrule
    \multicolumn{12}{l}{\textbf{Open-Source Models}} \\
    \midrule
    Qwen3-Omni             & \textbf{1498.92} & 5174.36 & 0.47 & \underline{0.25} & 0.02 & \underline{0.04} & \textbf{0.01} & \textbf{0.05} & \textbf{0.20}  & 4741.73 & \underline{6235.23} \\
    Qwen2.5-Omni           & 1317.74 & 5476.83 & 0.45 & \textbf{0.26} & 0.02 & 0.31 & 0.00 & \underline{0.04} & 0.17 & \underline{4540.06} & 7777.20 \\
    Phi-4-MM1              & 826.88 & 6462.43 & 0.33 & 0.08 & 0.01 & 0.15 & 0.00 & 0.01 & 0.08 & 6311.65 & 6831.08 \\
    Kimi-Audio             & 1336.20 & 5590.20 & 0.43 & 0.22 & 0.02 & 0.37 & 0.00 & \underline{0.04} & 0.18 & 4634.32 & 7928.59 \\
    Gemma-3n-E4B-it        & 1161.54 & 5815.46 & 0.41 & 0.17 & 0.01 & 0.05 & 0.00 & \underline{0.04} & 0.14 & 5402.73 & 6825.13 \\
    MiniCPM-o-2.6          & 1080.05 & 6600.83 & 0.44 & 0.22 & 0.02 & 0.54 & 0.00 & 0.01 & 0.14 & 5572.78 & 9115.74 \\
    Mimo-Audio             & 1445.39 & \textbf{4853.25} & \textbf{0.54} & 0.20 & \textbf{0.03} & 0.11 & 0.00 & 0.03 & 0.16 & \textbf{4113.93} & 6661.84 \\
    Mimo-Audio-Think       & \underline{1447.30}  & \underline{5008.01} & \underline{0.51} & 0.20 & \textbf{0.03} & \textbf{0.02} & 0.00 & \underline{0.04} & \underline{0.18} & 4599.73 & \textbf{6006.78} \\
    \bottomrule
  \end{tabular}
  \label{tab1}
\end{table*}

\subsection{Post Processing}
\label{post}

Following the above pipeline, we obtain AGL1K, a benchmark contains a broad range of audio categories including \emph{nature sounds}, \emph{animal vocalizations}, \emph{music}, and \emph{sounds of things} as well as many recordings that couple multiple categories and thus provide rich geo-informative clues. Geographically, the benchmark spans six continents and covers more than 72 countries and territories. 

After collecting ALM predictions, we observed that models often use aliases or informal variants for the same region (e.g., \emph{Oceania} being reported as \emph{Australia}, or \emph{United States} as \emph{USA}). To ensure consistent evaluation, we normalize model outputs to a unified location schema by mapping such variants to canonical region names, thereby improving the reliability of our benchmark results. 

In addition, we have developed an interactive audio geo-localization website, which can be deployed on Hugging Face Spaces to allow global users to explore and engage with the task. Please refer to Appendix~\ref{app:space}.

% 除此之外，我们还开发了一个audio geo-localization的网站，其可以部署在huggingface的space上让全世界的玩家体验，从而丰富audio geo-localization社区，具体请参见Appendix~\ref{app:space}.

\section{Experiments}

In this section, we investigate the following research questions:

\noindent \textbf{RQ1}: How do current audio-language models perform on the proposed benchmark?

\noindent \textbf{RQ2}: How do audio-language models infer potential geographic locations from acoustic clues?

\noindent \textbf{RQ3}: Do audio-language models exhibit geographic prediction inequality across regions?

\noindent \textbf{RQ4}: What are the primary causes of model prediction errors in audio geo-localization?

\begin{figure*}[t]
\begin{center}
\centerline{\includegraphics[width=2.1\columnwidth]{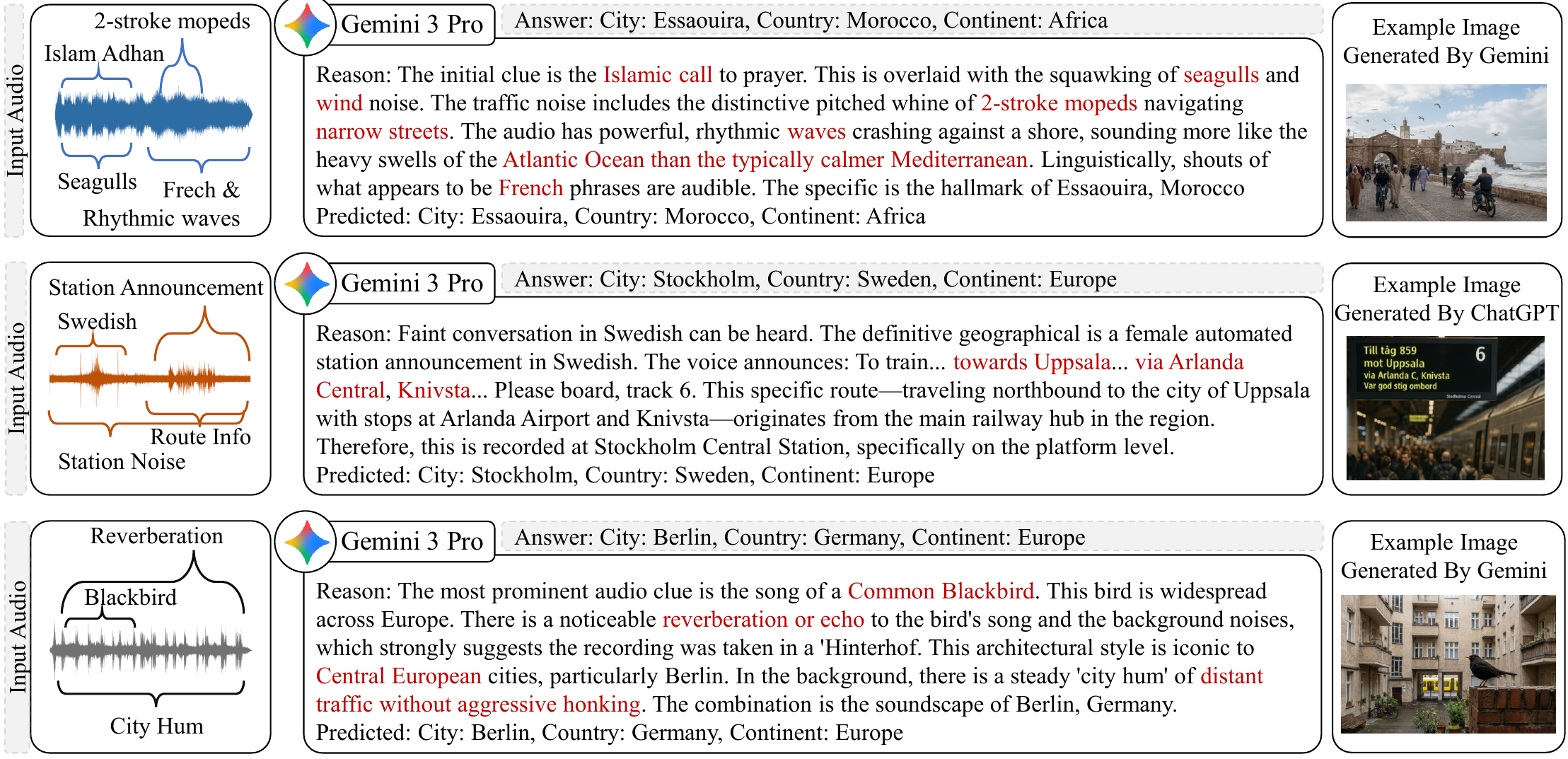}}
% 我们在这里选择了三段有代表性的音频，给出了他们音频clue的分布，Gemini 3 Pro推理的过程，并生成了一张图像以便理解。
\caption{\textbf{Benchmark examples.} We select three representative audio samples and present the distribution of their audio clues, the reasoning process of Gemini 3 Pro, and predictions of three other ALMs.}
\label{figure_case}
\end{center}
\end{figure*}

\subsection{Experimental Settings}

\subsubsection{Benchmarking Models} We benchmarked 8 state-of-the-art closed-source ALMs including: GPT-4o Audio Preview~\cite{gpt4o}, Gemini 3 Pro~\cite{gemini3}, Gemini 2.5 Pro~\cite{gemini25pro}, Gemini 2.5 Flash, Gemini 2.5 Flash-Lite, Gemini 2.0 Flash-Thinking~\cite{gemini2}, Gemini 2.0 Flash, Gemini 2.0 Flash-Lite. We also benchmarked 8 high-quality open-source ALMs including: Qwen3-Omni, Qwen2.5-Omni~\cite{qwen25omni}, Phi-4-MM1~\cite{phi4mm},  Kimi-Audio~\cite{Kimi-Audio}, Gemma-3n-E4B-it~\cite{gemma3}, MiniCPM-o-2.6~\cite{minicpmo}, Mimo-audio~\cite{mimoaudio}, and Mimo-audio-think~\cite{mimoaudio}. For open-source models except for Qwen3-Omni, we deploy them on a RTX 4090. Other models are accessed via their APIs with default settings.

\subsubsection{Evaluation Metrics} The metrics includes mean distance error(km) $\delta$, Geoscore based on Geoguessr game defined as $5000\exp(-\delta / 1492.7)$, hierarchical continent/country/city level accuracies, and thresholded accuracy at 1 km, 10 km, 500 km (denoted at $<*$) to capture both fine-grained and coarse localization performance. We also report a model’s reject rate (fraction of non-answers) as an indicator of robustness. Finally, to analyze reliance on speech clues, we compute average distance errors separately for speech and non-speech audio segments. The definition of each metric can be found in the Appendix~\ref{app:met}.

\subsection{Comparison Results (RQ1)}

% 大模型涌现出了一定的Audio GeoLocalization能力
\textbf{Modern audio-language models have begun to exhibit emergent capabilities in audio geo-localization.}
Table~\ref{tab1} reports the performance of all models on the Audio Geo-Localization Benchmark. It shows Gemini 3 Pro achieves the strongest overall results, with 19\% of examples localized within 10 km, continent-level accuracy of 0.82, and country-level accuracy of 0.51. However, a pronounced gap exists between closed-source and open-source models: the best open-source model, Mimo-audio, records an average error of 4853 km. It is more than twice that of Gemini 3 Pro.

% Audio GeoLocalization能力是随着模型scaling而scaling的
\textbf{Audio geo-localization performance scales systematically with model capacity.} Within the Gemini family, we observe a clear advance. Distance error decreases from 2992 km in Gemini 2.0 Flash-Thinking to 2181 km in Gemini 3 Pro, alongside steady gains in various accuracies. Notably, some Gemini 2.0 Flash variants outperform Gemini 2.5 Flash, implying that the distilled model may prioritize efficiency or task-specific objectives that do not directly align with audio geo-localization. However, reasoning-enhanced variants (e.g., “thinking” modes) do not yield consistent improvements, suggesting that explicit reasoning alone provides limited benefit.

% 语言是明显的一个声音定位的脚手架。Thinking对于定位帮助不大，甚至会有反作用。
\textbf{Linguistic content serves as a dominant scaffold for audio-based geo-localization}. For Gemini 3 Pro, the average localization error drops from 3727 km on non-speech audio to \textbf{1548 km} when speech is present. Similar gaps appear across all evaluated audio-language models from closed-source to open-source models. This result highlights the dominant role of linguistic clues, while purely environmental sounds remain more challenging. 

Additional statistics and comparisons are provided in Appendix~\ref{app:comp}.

\subsection{Benchmark Examples (RQ2)} 

To qualitatively examine the reasoning behavior of modern audio models, we analyze Gemini 3 Pro's reasoning output across three representative scenarios, shown in Figure~\ref{figure_case}, along with the prediction of GPT-4o Audio, Qwen3-Omni, and Mimo-Audio. The complete reasoning traces are provided in the Appendix~\ref{app:out}.

The first case requires integrating linguistic and environmental signals. In addition to recognizing Islamic Adhan, Gemini 3 Pro detects coastal wind, seagull calls, French Speech, and motorcycle noise in narrow streets, jointly localizing the recording to Essaouira, Morocco. However, most other models overemphasize Adhan and lead to predictions of Istanbul or Mumbai.

The second case tests geographic inference from indirect clues. The audio references travel toward Uppsala via Arlanda and Knivsta, leveraging regional transit knowledge. Gemini 3 Pro, GPT-4o, and Mimo-Audio correctly infer Stockholm as the recording location.

This third case contains no spoken language. Instead, Gemini 3 Pro relies on non-verbal clues, including the call of a Common Blackbird, architectural style, and background traffic patterns, to correctly localize the scene to Berlin, Germany. Qwen3 and Mimo, however, identified the European bird and randomly guessed a place in Europe without further thinking.

Together, these case studies show that \textbf{leading audio models can exploit linguistic, environmental, and contextual clues to perform geographic reasoning}, and \textbf{our benchmark captures these compositional reasoning modes in a unified evaluation framework.}

\begin{figure*}[t]
\begin{center}
\centerline{\includegraphics[width=2\columnwidth]{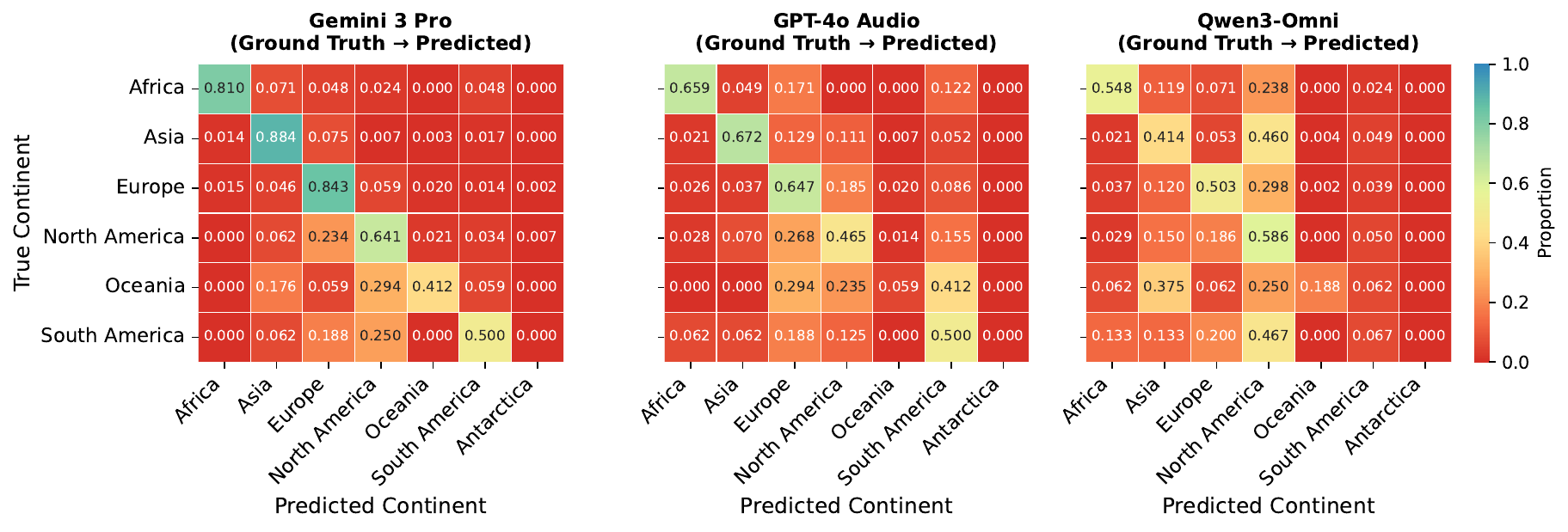}}
\caption{The continent-level prediction inequality in audio geo-localization. The i-j entry indicates that the truth is the portion of the continent in the i-th line that is predicted to be continent in the j-th row.}
\label{figure_inequal}
\end{center}
\end{figure*}

\subsection{Continent-Level Prediction Inequality (RQ3)} 

% 为了检验模型预测的地区是否有倾向性，Figure \ref{figure_inequal} reveals pronounced continent-level disparities in audio geo-localization performance across models. A key finding is most models fails almost entirely on Oceania, systematically misclassifying Oceania samples as Europe or North America, whereas Gemini 3 Pro correctly assigns a substantial fraction of Oceania samples to the correct class. However, this improvement is accompanied by a new failure mode: Gemini 3 Pro occasionally predicts Antarctica, an implausible outcome that is largely absent in earlier models.

% We also observe strong regional asymmetry across models. Mimo-audio shows comparatively high accuracy on Asia, exceeding GPT-4o Audio on Asian samples, while underperforming on several other continents. This pattern suggests Mimo, as a model trained by Chinese company, potentially exposed to more Asian audio during training. 总而言之，不同的模型均存在着一定程度的地理Inequality，这可能是未来改进Audio Localizability的核心。

Figure 6 shows continent-level confusion (row-normalized). Gemini 3 Pro has the strongest diagonal, meaning the most stable continent recognition, especially for Africa/Asia/Europe and performance drops on Oceania/South America.
GPT-4o Audio is less concentrated on the diagonal. It makes broader cross-continent mistakes, and Oceania is particularly unstable, often being predicted as other continents.
Qwen3-Omni shows clear label collapse toward North America: many samples from other continents are predicted as North America, leading to poor results on Oceania and South America.
Overall, these results indicate that \textbf{geographic prediction inequality is a persistent issue across audio-language models, highlighting the need to explicitly account for regional fairness in future work}.

\subsection{Error Distribution (RQ4)} 
\begin{figure}[h]
\begin{center}
\centerline{\includegraphics[width=\columnwidth]{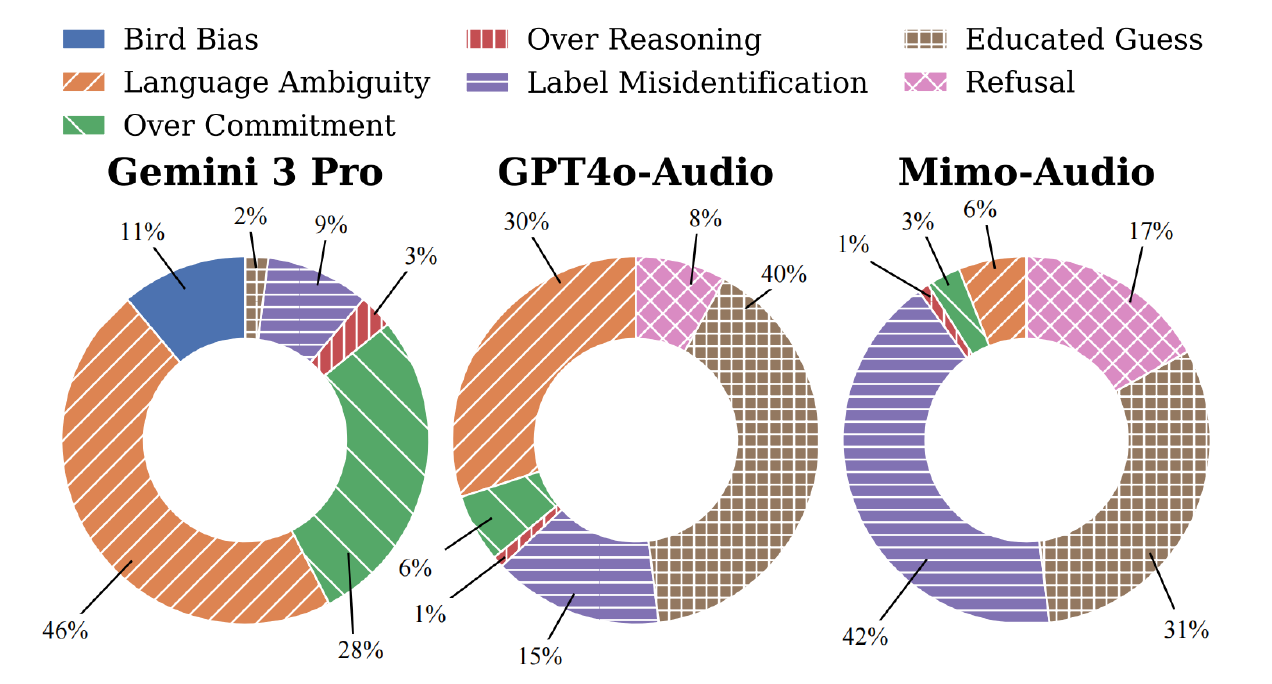}}
\caption{\textbf{Error distribution across three models.}}
\label{fig_error}
\end{center}
\end{figure}

To analyze model failure causes, we annotate 300 country-level misclassifications for three representative models. Errors are grouped into seven categories: Bird Bias, Language Ambiguity, Over-Commitment, Over-Reasoning, Label Misidentification, Educated Guess, and Refusal, capturing both perceptual and reasoning-related failures. The definition of each error is elaborated in the Appendix~\ref{app:err}.

Figure~\ref{fig_error} shows that error distributions differ substantially across models. For Gemini 3 Pro, Language Ambiguity dominates, followed by Over-Commitment and Bird Bias. This reflects a tendency to exploit fine-grained acoustic clues, particularly language and bird vocalizations, which improves localization accuracy but also introduces distinctive errors. In contrast, GPT-4o Audio is primarily affected by Educated Guess and Language Ambiguity, indicating frequent fallback to uncertain predictions when informative clues are weak. Mimo-Audio exhibits a different pattern, with Label Misidentification accounting for the largest share of errors, followed by Educated Guess and Refusal, suggesting limitations in multilingual audio discrimination and confidence calibration.

% 总而言之，本实验给出各类模型提升Audio Localization的能力的可能方案：在基础上，优化多语种识别的能力；在决策过程中，主动进行深度推理，尽量避免educated guess。在标签分析上，避免过于依赖单一标签，如直接基于语言和鸟类下判断。

In summary, this experiment suggests several actionable directions for improving audio geo-localization across model families. At the perception level, strengthening multilingual speech recognition is essential for reducing language-related errors. At the reasoning level, models should adopt more deliberate and evidence-aware reasoning to minimize reliance on uninformed educated guesses. At the fusion level, models should avoid over-dependence on single clues, such as language or bird vocalizations, and instead integrate multiple complementary audio signals. Together, \textbf{these findings provide practical guidance for enhancing the accuracy and robustness of audio-based localization systems.}

\section{Conclusion}

% \item \textbf{The first ALM-based benchmark for audio geo-localization.} We introduce AGL1K, a benchmark comprising 1,444 user-uploaded audio clips from 72 countries and territories across six continents, covering diverse acoustic scenes including natural soundscapes, animal vocalizations, music, human-made sounds, and spoken conversations.

% \item \textbf{A principled notion of Audio Localizability.} We define and operationalize Audio Localizability, a quantitative measure of an audio clip's geographic informativeness. The metric aggregates temporally localized evidence from predicted sound categories during inference, enabling filtering and analysis of localizable recordings.

% \item \textbf{A comprehensive evaluation of state-of-the-art ALMs.} We benchmark 16 representative ALMs and show that audio geo-localization is tractable yet remains challenging. We further analyze models' reasoning traces, regional biases, error modes, and the interpretability of the proposed localizability metric.

We introduce AGL1K, a benchmark for audio geo-localization. Built on user-uploaded recordings and filtered by our proposed audio localizability, the benchmark comprises 1,444 audio clips from 72 countries and territories across six continents, covering diverse acoustic scenes. Through the evaluation of 16 ALMs, we show that geo-localization ability has emerged in ALMs while AGL1K still remains challenging. Our analyses identify major error modes and suggest directions for improving ALMs, including strengthening fine-grained perception, reducing regional bias, and enhancing compositional reasoning.
As audio geo-localization requires compositional reasoning on audio clues, we expect audio geo-localization to emerge as an important benchmark for evaluating ALMs' reasoning capability.

\section{Limitations}

% 由于原始数据分布的原因，各大洲的数据分布不太均衡。

\begin{itemize}
    
\item Due to the inherent distribution of the original data, the number of samples is uneven across continents. In particular, Europe, Asia, and North America are significantly overrepresented compared to other regions such as Africa, Oceania, and South America. 

\item The diversity of sound labels, as well as the distribution of positive and negative attribution results, is partially constrained by the choice of annotation models and the collected data. 

\end{itemize}

% The current benchmark setup does not support the integration of external tools. Enhancing the setup with access to auxiliary systems such as retrieval engines or speech recognition could potentially improve the quality of model outputs.

% Subjective bias is inevitable during the annotation of contribution types, as human preferences and interpretations may vary. 

% Bibliography entries for the entire Anthology, followed by custom entries
%\bibliography{anthology,custom}
% Custom bibliography entries only
\bibliography{custom}

\appendix

\newpage 
\section{Appendix}

\subsection{Introduction of Aporee Platform}
\label{apore}

Aporee, launched in 2006, provides geo-tagged audios by enabling users to voluntarily upload environmental sound recordings with associated GPS coordinates, resulting in a global sound map. The platform covers a wide range of acoustic environments, including urban areas, rural regions, and natural landscapes, and has been widely used for artistic, educational, and research purposes. Through collaboration with the Aporee team, we obtained access to their audio data under the Apache license.

\subsection{Interactive Platform Construction}
\label{app:space}

\begin{figure}[h]
\begin{center}
\centerline{\includegraphics[width=\columnwidth]{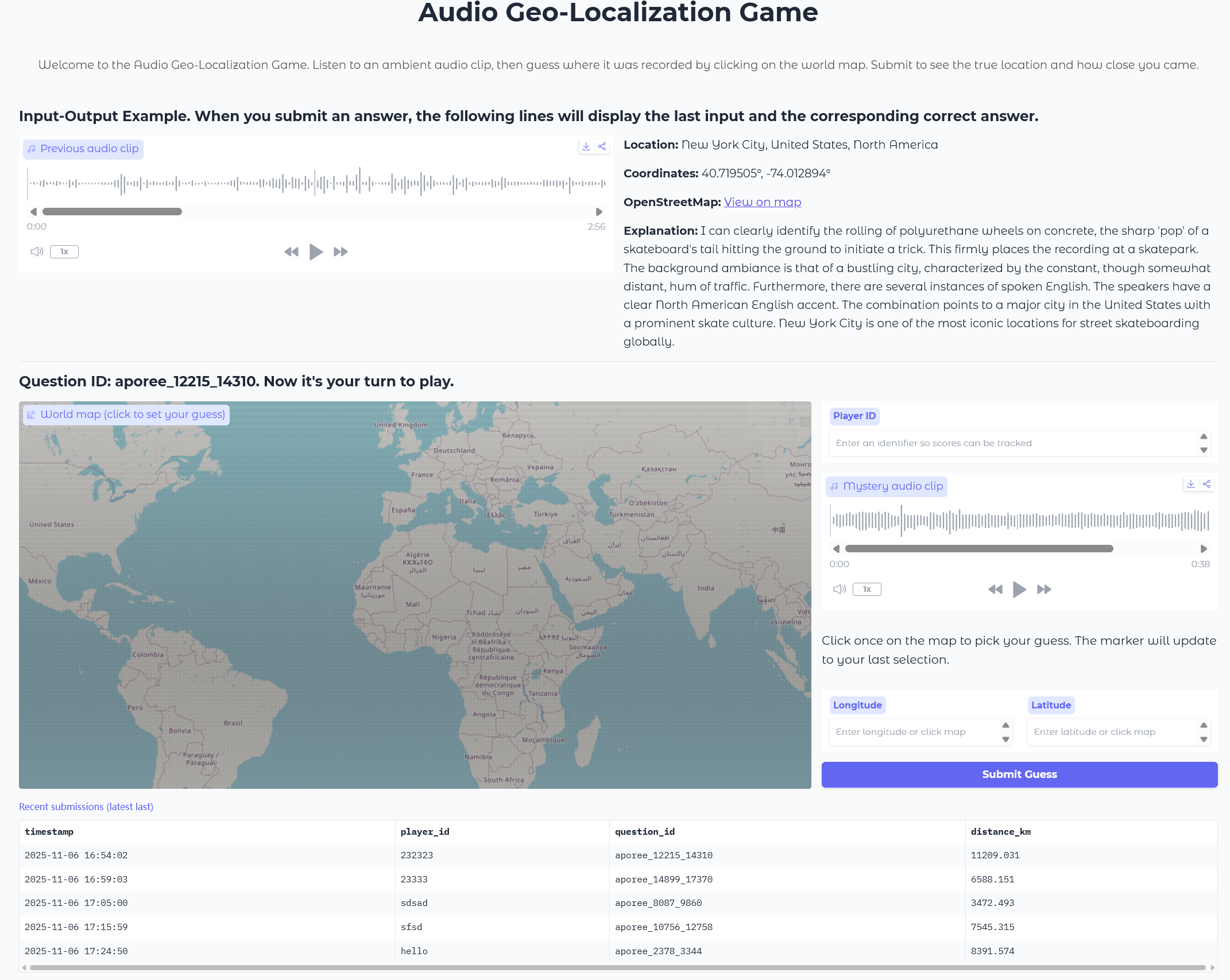}}
\caption{\textbf{The screenshot of our Interactive Platform}. Built by Hugging Face Gradio, our platform can be easily deployed to Hugging Face Space or locally.}
\label{figure4}
\end{center}
\end{figure}

To collect human performance data and encourage broader participation in audio geo-localization, we develop an interactive web-based platform and deploy it on HuggingFace. The interface presents users with an audio clip and an interactive map, and users are asked to infer the sound’s location by clicking on the map. User predictions and anonymized identifiers are publicly logged on the HuggingFace Space, enabling analysis of human localization behavior and providing a reference baseline for comparison with model performance.

\subsection{Formulation of the four acoustic filters}
\label{app:fils}

The four acoustic filters are defined as:

\begin{equation}
\begin{aligned}
\text{RMS} &= \sqrt{\frac{1}{N} \sum_{n=1}^{N} x[n]^2}, \\ 
\text{SF} &= \frac{\left(\prod_{k=1}^{K} P[k]\right)^{1/K}}{\frac{1}{K} \sum_{k=1}^{K} P[k]}, \\
\text{CR} &= \frac{\#\{n \mid |x[n]| = x_{\max}\}}{N}, \\ 
\text{AC} &= \sum_{f} \frac{\sum_{t=1}^{T-1} |E_{f,t+1} - E_{f,t}|}{\sum_{t=1}^{T} E_{f,t}}, \\ 
\end{aligned}
\end{equation}

where $x[n]$ denotes the audio amplitude at sample $n$, $P[k]$ the power spectral density at frequency bin $k$, and $E_{f,t}$ the energy of frequency band $f$ at time frame $t$.

\begin{table*}[t]
  \centering
  \small
  \setlength{\tabcolsep}{3pt} % 调小列间距，适配两栏
  \caption{Further Comparison Results.}
  \label{tab:futrthur}
  \begin{tabular}{lrrrrrrr}
    \toprule
    Model &
\makecell{q25$\downarrow$} &
\makecell{q50$\downarrow$} &
\makecell{q75$\downarrow$} &
\makecell{Animal\\Distance$\downarrow$} &
\makecell{Music\\Distance$\downarrow$} &
\makecell{Nature\\Distance$\downarrow$} &
\makecell{Things\\Distance$\downarrow$} \\
    \midrule
     RANDOM & 6253.81 & 10062.38 & 13639.9 & 10181.77 & 10667.35 & 10259.51 & 10515.23 \\ 
    \midrule
    \multicolumn{8}{l}{\textbf{Closed-Source Models}} \\
    \midrule
     Gemini 3 Pro & \textbf{55} & \textbf{480.19} & \textbf{1888.6} & \textbf{3143.16 }& \textbf{3231.5} & \textbf{3669.51} & \textbf{3765.42} \\ 
    Gemini 2.5 Pro & 97.59 & 548.1 & 2848.37 & 4635.67 & 3476.44 & 4620.69 & 4541.46 \\ 
    Gemini 2.5 Flash & 279.21 & 1213.16 & 6782.78 & 5962.53 & 3893.8 & 5918.76 & 5353.91 \\ 
    Gemini 2.5 Flash-Lite & 664.15 & 2295.73 & 8275.64 & 5976.78 & 4092.25 & 6247.15 & 5911.77 \\ 
    Gemini 2.0 Flash Thinking & 263.39 & 883.79 & 5544.65 & 4576.98 & 4889.07 & 4839.28 & 4808.03 \\ 
    Gemini 2.0 Flash & 218.81 & 886.77 & 5119.63 & 4510.66 & 4319.07 & 4673.66 & 4462.81 \\ 
    Gemini 2.0 Flash-Lite & 280.3 & 1002.83 & 5888.17 & 5936.17 & 3900.27 & 5883.09 & 5426.24 \\ 
    GPT-4o Audio Preview & 415.21 & 1715.7 & 7723.63 & 7960.98 & 4849.44 & 7696.62 & 6613.44 \\ 
 
    \midrule
    \multicolumn{8}{l}{\textbf{Open-Source Models}} \\
    \midrule
    Qwen3-Omni & \textbf{633.98} & 5627.17 & \textbf{8821.64} & \textbf{5920.09} & 5048.03 & 6064.08 & 6484.72 \\ 
    Qwen2.5-Omni & 983.09 & 5952.77 & 10000 & 7723.77 & 5694.41 & 7871.21 & 7854.18 \\ 
    Phi-4-MM1 & 1797.82 & 6485.13 & 10000 & 6905.37 & 6066.34 & 6728.47 & 6640.15 \\ 
    Kimi-Audio & 829.04 & 5945.86 & 10000 & 8274.65 & 6692.15 & 8148.38 & 7601.32 \\ 
    gemma-3n-E4B-it & 1263.42 & 7020.47 & 9441.25 & 7203.56 & 6063.93 & 7160.14 & 6794.96 \\ 
    MiniCPM-o-2.6 & 1355.05 & 10000 & 10000 & 9517.41 & 7424.07 & 9345.83 & 9070.55 \\ 
    Mimo-Audio & 893.83 & \textbf{3450.54} & 9208.35 & 7318.02 & 4552 & 7293.23 & 6091.1 \\ 
    Mimo-Audio-think & 854.09 & 3627.95 & 9036.6 & 6046.2 & \textbf{4419.81} & \textbf{6028.08} & \textbf{6001.72} \\ 
    \bottomrule
  \end{tabular}
  \label{tab1}
\end{table*}

\subsection{Definition of Evaluation Metrics}

\label{app:met}

Let $\mathcal{D}={(x_i, y_i)}_{i=1}^{N}$ denote the evaluation dataset, where $x_i$ is an audio sample and $y_i=(c_i^{\text{cont}}, c_i^{\text{country}}, c_i^{\text{city}}, \ell_i)$ denotes its ground-truth continent, country, city, and geographic coordinates $\ell_i=(\phi_i,\lambda_i)$ (latitude and longitude).
For a model prediction $\hat{y}_i=(\hat{c}_i^{\text{cont}}, \hat{c}_i^{\text{country}}, \hat{c}_i^{\text{city}}, \hat{\ell}_i)$, we define the following metrics.

\subsubsection{Mean Distance Error (km)}

We compute the great-circle distance between predicted and ground-truth coordinates using the Haversine formula:
\begin{equation}
d_i = \mathrm{Haversine}(\ell_i, \hat{\ell}*i).
\end{equation}
If the model refuses to answer, we directly set $d_i$ as $10000$. The mean distance error is
\begin{equation}
\text{DistErr} = \frac{1}{|\mathcal{A}|} \sum*{i\in \mathcal{A}} d_i,
\end{equation}
where $\mathcal{A}\subseteq{1,\dots,N}$ denotes the set of all samples. 

\subsubsection{Hierarchical Accuracy}

We report accuracy at three geographic levels:
\begin{equation}
\begin{aligned}    
\text{Acc}_{L} = \frac{1}{|\mathcal{A}|} \sum_{i\in \mathcal{A}} \mathbb{I}\big[\hat{c}_i^{L} = c_i^{L}\big], \\
\quad L \in {\text{continent}, \text{country}, \text{city}},
\end{aligned}
\end{equation}
where $\mathbb{I}[\cdot]$ is the indicator function.

\subsubsection{Thresholded Distance Accuracy ($<\tau$)}

To capture coarse-to-fine localization performance, we compute thresholded accuracy at multiple distance thresholds $\tau \in {1,10,100,1000}$ km:
\begin{equation}
\text{Acc}_{<\tau} = \frac{1}{|\mathcal{A}|} \sum_{i\in \mathcal{A}} \mathbb{I}\big[d_i < \tau \big].
\end{equation}
These correspond to the columns denoted as $<1$, $<10$, $<100$, and $<1000$ in Table~\ref{tab1}.

\subsubsection{Reject Rate}

Some models may abstain from answering. We define the reject rate as
\begin{equation}
\text{RejectRate} = 1 - \frac{|\mathcal{A}_a|}{N},
\end{equation}
which measures the fraction of samples for which the model does not return a prediction and $\mathcal{A}_a$ denotes the set of samples for which the model produces a valid prediction (i.e., not rejected).

\subsubsection{Speech and Non-Speech Distance Error}

To analyze reliance on linguistic clues, we partition the dataset into speech samples $\mathcal{D}*{\text{speech}}$ and non-speech samples $\mathcal{D}_{\text{non-speech}}$.
We then compute mean distance error separately:
\begin{equation}
\begin{aligned}
\text{DistErr}_{\text{speech}} = & \frac{1}{|\mathcal{A}_{\text{speech}}|} \sum_{i\in \mathcal{A}_{\text{speech}}} d_i,
\\
\text{DistErr}_{\text{non-speech}} =& \frac{1}{|\mathcal{A}_{\text{non-speech}}|} \sum_{i\in \mathcal{A}_{\text{non-speech}}} d_i.
\end{aligned}
\end{equation}

\subsection{Definition of Main Sound Labels}

\textbf{Nature} refers to the Natural Sound label in Audioset. This is defined as: Sounds produced by natural sources in their normal soundscape, excluding animal and human sounds.

\textbf{Animal} refers to the Animal label in Audioset. This is defined as: All sound produced by the bodies and actions of nonhuman animals.

\textbf{Music} refers to the Music label in Audioset. This is defined as: Music is an art form and cultural activity whose medium is sound and silence. The common elements of music are pitch, rhythm, dynamics, and the sonic qualities of timbre and texture.

\textbf{Things} refers to the Sounds of Things label in Audioset. This is defined as: Set of sound classes referring to sounds that are immediately understood by listeners as arising from specific objects (rather than being heard more literally as "sounds").

\begin{figure*}[t]
\begin{center}
\centerline{\includegraphics[width=2\columnwidth]{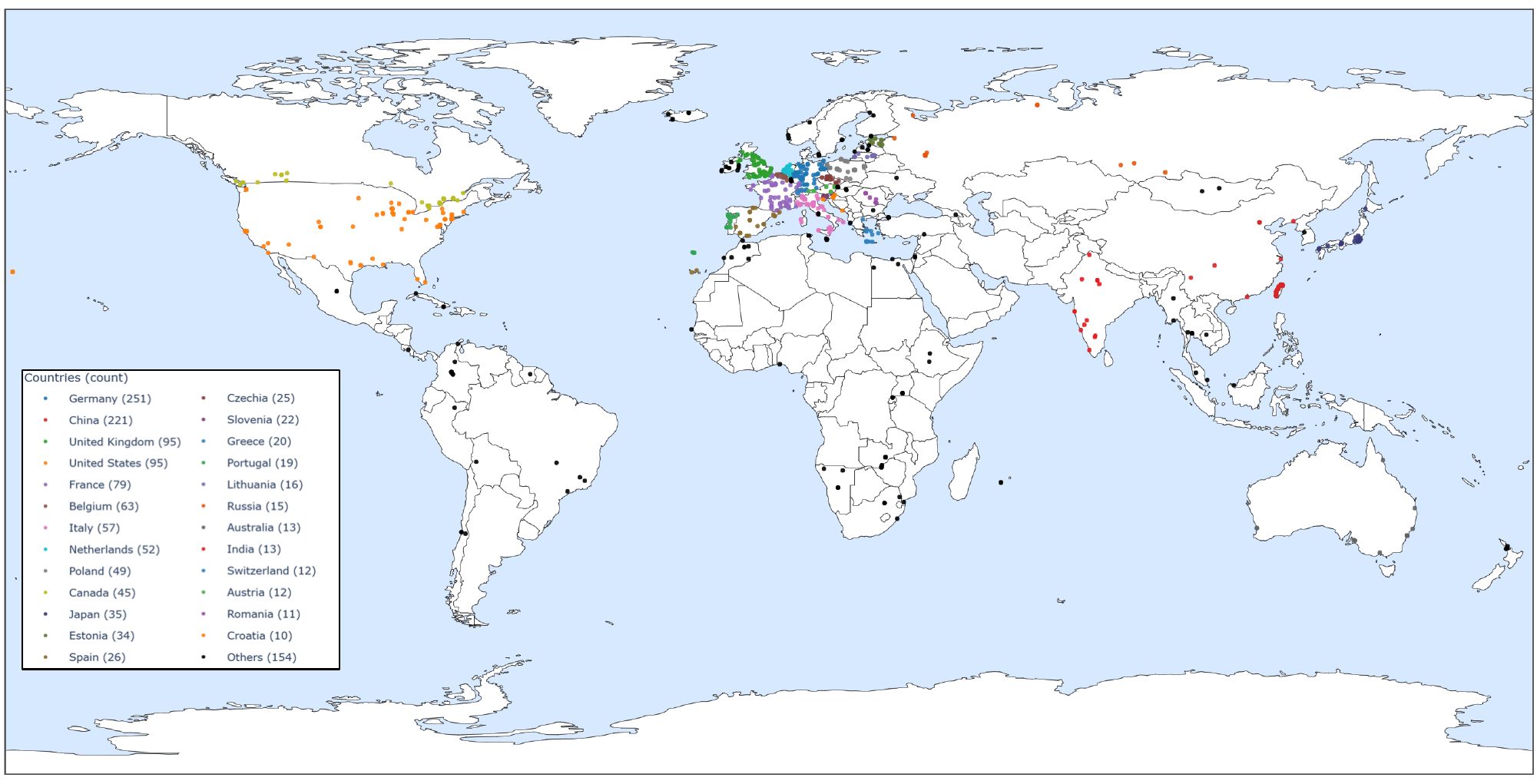}}
\caption{\textbf{The Global Distribution of AGL1K.} Locations of all 1,444 audios across 74 countries.
Each marker denotes an image coordinate, and colors indicate country (top contributors listed; “Others” aggregates
the remainder)}
\label{figure_global}
\end{center}
\end{figure*}

\subsection{Further Comparison Results}
\label{app:comp}
% 我们在这里提供了各个模型的1/4分位点，1/2分位点，以及3/4分位点的误差数据，以q25，q50，q75代表。我们还体哦概念股了，Animal，Music，Nature，Sound of Things各个子类别下的误差数据。可以看到Gemini3 Pro在25的数据上，都有小于50公里的误差，证明了其惊人的声音定位能力，而开源模型最好的仍有634km，体现了开源模型和闭源模型的鸿沟。而由于Music可能能够提供一定的语音线索，Animal的叫声也能通过分布进行判断，使得这两项误差较Nature和Sound of Things更低。

We report the 25th, 50th, and 75th percentile localization errors (denoted as q25, q50, and q75, respectively) for each model in Table~\ref{tab:futrthur}. Additionally, we break down the errors across four subcategories: Animal, Music, Nature, and Sound of Things. Notably, Gemini 3 Pro achieves localization errors under 50 km at the 25th percentile across all categories, highlighting its remarkable audio grounding capabilities. In contrast, the best-performing open-source model still exhibits an error of 634 km at the same percentile, underscoring the significant performance gap between proprietary and open-source models. The relatively lower errors in Music and Animal categories may be attributed to the presence of linguistic or distributional cues in musical content and animal vocalizations, whereas Nature and Sound of Things categories tend to lack such informative structure.

\subsection{Further Experimental Settings}
% 我们使用Openrouter(\url{https://openrouter.ai/models})调用了closed-source models以及阿里云bailian{\url{https://cn.aliyun.com/product/bailian}}的API调用了Qwen3-Omni。具体价格请参考对应页面。开源模型部署在一张4090上，每个模型花费时间均约为12GPU Hour。

We accessed closed-source models via OpenRouter (\url{https://openrouter.ai/models}
) and invoked Qwen3-Omni through Alibaba Cloud’s Bailian API (\url{https://cn.aliyun.com/product/bailian}
). Please refer to the respective websites for detailed pricing information. Open-source models were deployed on a single NVIDIA RTX 4090 GPU, with each model requiring approximately 12 GPU hours for inference.

\section{Definition of Each Error Type}
\label{app:err}

\textbf{Bird Bias} denotes errors arising from region-specific bird vocalizations. 

\textbf{Language Ambiguity} refers to cases in which a language is spoken across multiple geographic regions, leading the model to select an incorrect location despite correctly identifying the language. 

\textbf{Over-Commitment} captures situations in which the model’s intermediate reasoning supports multiple plausible hypotheses, but it nevertheless commits to a single overconfident prediction. 

\textbf{Over-Reasoning} describes cases in which the model initially infers the correct answer but subsequently overrides it due to unnecessary or misguided additional reasoning. Label Misidentification corresponds to factual inference errors, most commonly incorrect language identification (e.g., classifying French as German). 

\textbf{Educated Guess} refers to instances in which the model reports insufficient information and effectively responds at random. 

\textbf{Refusal} covers cases in which the model declines to provide an answer.

\subsection{Global Distribution of AGL1K}

The Figure~\ref{figure_global} shows the global distribution of AGL1K.

\subsection{Consistency on three contribution-labeling LLM} 
\label{app:consi}

\begin{figure}[h]
\begin{center}
\centerline{\includegraphics[width=\columnwidth]{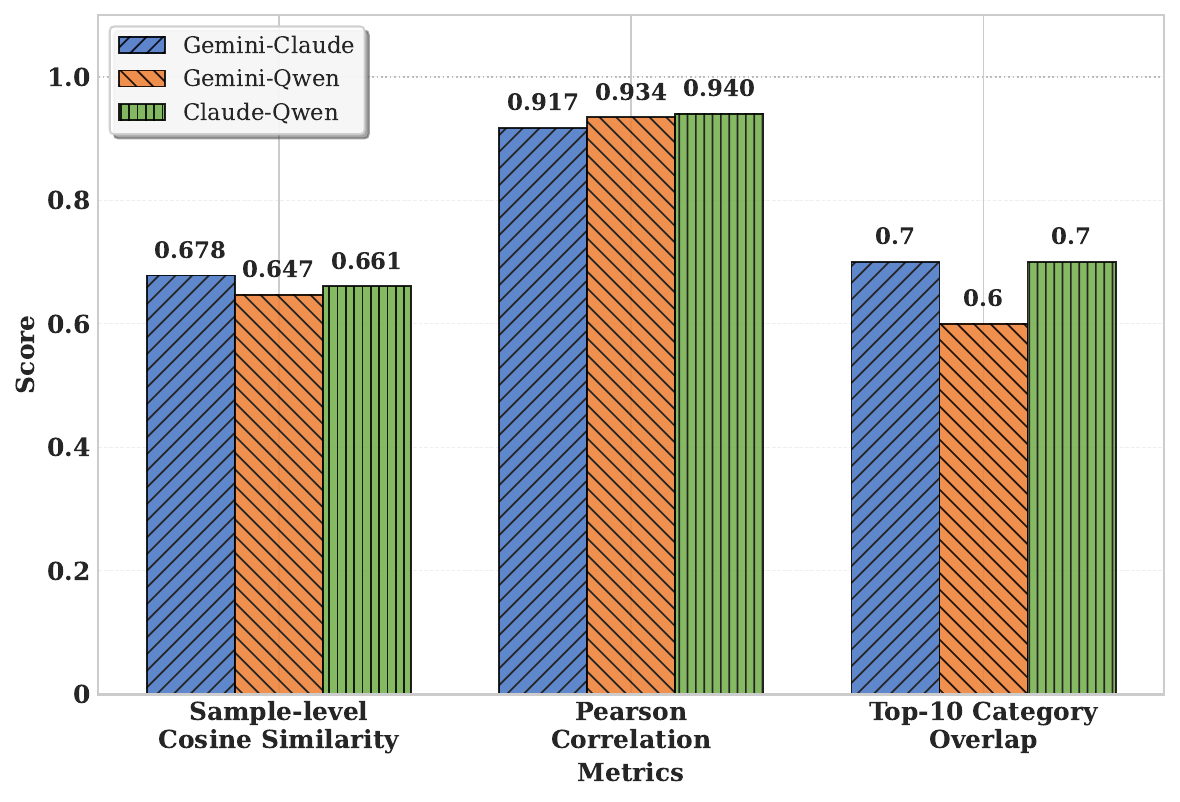}}
\caption{Model Similarity Comparison Across Different Metrics.}
\label{fig_sim}
\end{center}
\end{figure}
% 我们首先对每个 question_id，计算三个模型 contribution 向量的两两余弦相似度，发现三个模型在样本级别相似度中等偏高（0.65-0.68），说明对同一音频的贡献度评估方向相似。接着，我们通过汇总所有类别的 contribution 值后，计算三个模型之间的相关性，发现PCC高达0.9，说明整体上三个模型的 contribution 分布高度一致。最后我们计算三个模型 Top-10 贡献度类别的重叠程度，也发现了三个模型在最重要的类别上有一半以上重叠。这说明三个模型的结果具有一定的一致性，通过使用三个模型进行label，能够取得稳定但对不同部分仍有侧重contribution labeling的结果。
To assess the reliability and instruction-following consistency of large language models in contribution evaluation, we measure cross-model agreement among Gemini, Claude, and Qwen using three complementary metrics (Figure~\ref{fig_sim}). At the sample level, pairwise cosine similarity of contribution vectors ranges from \textbf{0.65 to 0.68}, indicating substantial directional agreement on which audio clues matter for individual examples. When aggregating contributions across categories, the Pearson correlation exceeds \textbf{0.91} for all model pairs, demonstrating highly consistent global attribution patterns. In addition, the Top-10 category overlap remains above \textbf{0.60}, showing that models agree on the majority of the most influential sound categories.

Together, these results indicate that contribution judgments are stable across different LLMs, supporting the robustness of our ensemble-based annotation strategy while allowing for minor model-specific preferences.

\subsection{Continent-Distribution of Localizability}

\begin{figure}[h]
\begin{center}
\centerline{\includegraphics[width=\columnwidth]{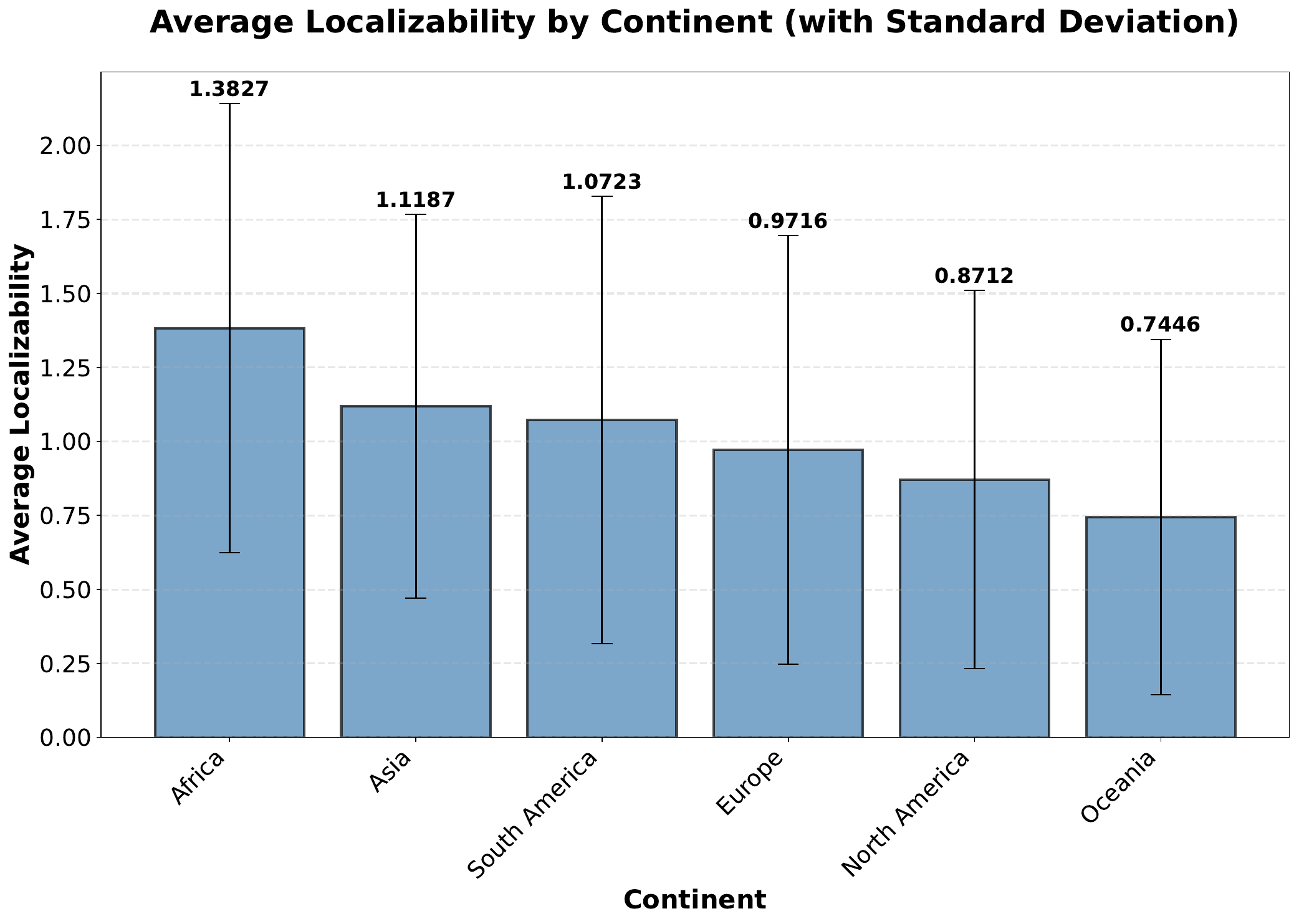}}
\caption{Continent-Distribution of Localizability.}
\label{fig_con_dis}
\end{center}
\end{figure}

% 我们分析了不同大洲的可定位性。如图~\ref{fig_con_dis}所示。我们发现Africa Asia South America有着较高的localizability，而Europe North America Oceania则较低。这可能潜在上与语言相关，前三者具有较为特色的文化与语言，而后三者可能很多地区都是英语国家，使得可定位性有一定的下降。此外，该可定位性的顺序与误差也有一定的相关性，gemini3 Pro预测准确率的排序是Asia，Europe，Africa，North America， South America和Oceania，说明了我们的localizability能一定程度上与模型的预测误差相关。

We analyze the localizability of audio across different continents, as shown in Figure~\ref{fig_con_dis}. We observe that Africa, Asia, and South America exhibit higher average localizability scores, whereas Europe, North America, and Oceania show lower values. One possible explanation lies in the linguistic and cultural distinctiveness of the former group: regions within Africa, Asia, and South America often feature more unique acoustic environments and languages, which may provide richer cues for geo-localization. In contrast, countries in Europe, North America, and Oceania often share English as a dominant language, potentially leading to less discriminative audio patterns and thus reduced localizability.

Moreover, the observed localizability rankings partially align with the actual model performance. For instance, Gemini 3 Pro achieves the highest prediction accuracy in Asia, followed by Europe, Africa, North America, South America, and Oceania. which correlates with the average localizability scores across continents. This suggests that localizability, as a metric, captures intrinsic properties of audio that influence model performance and may serve as a useful predictor of geo-localization difficulty across regions.

\subsection{Human Annotator}

This study involved human annotators at two key stages. First, during dataset refinement, after filtering about 3,000 samples using the localizability metric, a human annotator was tasked with selecting examples that were genuinely geo-locatable. The annotator, a PhD student in computer science, was instructed to consider the model's output, the audio content, and the ground-truth location to assess whether city-level inference was plausibly achievable.

Second, in the error analysis stage, another human annotator was responsible for labeling failure cases according to predefined error categories. For speech-related errors, the annotator consulted a PhD student in linguistics to determine whether language recognition was incorrect. Annotators were compensated at a rate of 100 RMB per hour and were informed that their annotations would be used in this study.

\subsection{Prompts}

\subsubsection{Benchmark Prompt}

\begin{tcolorbox}[colback=blue!3!white,
                  colframe=blue!75!black,
                  title=Benchmark Prompt,
                  fonttitle=\bfseries,
                  coltitle=white,
                  sharp corners=south,
                  breakable]

You are an expert audio analyst with exceptional ability to identify geographical locations from environmental sounds. Listen carefully to the provided audio recording and identify the geographical location where it was recorded.

You MUST provide your best guess regardless of uncertainty. Even if you feel unsure, you must still make an educated guess based on the audio characteristics, ambient sounds, languages spoken, cultural indicators, or any other audio clues you can detect.

Your response MUST be in the following JSON format (replace the values with your predictions):

\{

    \indent "reason": "Detailed explanation of your reasoning process. Describe what audio clues you heard (e.g., language, accent, background sounds, environmental noise, cultural indicators, weather sounds, traffic patterns, animal sounds, etc.) and how they led you to this specific location prediction. Be specific about the evidence that supports your geographical conclusion."
    
    \indent "city": "Your predicted city name",
    
    \indent "country": "Your predicted country name", 
    
    \indent "continent": "Your predicted continent name",
    
    \indent "longitude": predicted longitude value,
    
    \indent "latitude": predicted latitude value
    
\}

Important requirements:
\begin{itemize} 
\item 1. You MUST respond with a complete JSON object - no explanations outside the JSON, no refusals, no uncertainty statements
\item 2. If you're unsure about specific coordinates, provide your best numerical estimate
\item 3. All string values should be in English
\item 4. Longitude should be between -180 and 180
\item 5. Latitude should be between -90 and 90
\item 6. The "reason" field must contain your detailed step-by-step analysis of the audio clues
\item 7. Even if the audio seems generic, you must still provide a location guess with reasoning
\end{itemize}

Listen to the audio and provide your JSON response now:
\end{tcolorbox}

\subsubsection{Contribution Calculation Prompts}

\begin{tcolorbox}[colback=blue!3!white,
                  colframe=blue!75!black,
                  title=Contribution Calculation Prompts,
                  fonttitle=\bfseries,
                  coltitle=white,
                  sharp corners=south,
                  breakable]

You are analyzing an audio geolocalization reasoning text. Your task is to identify which sound categories from the provided list contributed to the location prediction, and assess their contribution level (0.0 to 1.0).

Available sound categories:

\{categories\}

Please analyze the reasoning text and identify which categories are mentioned or implied, and how much each category contributed to the location prediction. 

Output your response as a JSON object where:

- Each key is a category name (exactly as listed above)

- Each value is a contribution score between 0.0 and 1.0, where:

  * 0.0 = not mentioned/not relevant
  
  * 0.1-0.3 = mentioned but minor contribution
  
  * 0.4-0.6 = moderate contribution
  
  * 0.7-0.9 = significant contribution
  
  * 1.0 = primary/dominant contribution

Only include categories that have a contribution > 0.0. Output ONLY the JSON object, no additional text or explanation.

Example output format:

\{

  "Speech": 0.8,
  
  "Music": 0.3,
  
  "Animal": 0.2

\}
\end{tcolorbox}

\subsection{Sample Output}
\label{app:out}
In this section, we present the complete outputs of selected models for three representative cases. For each audio clip, we additionally generate a corresponding image using NanoBanana Pro to facilitate the interpretation and understanding of the audio.

\subsubsection{Example 1-Gemini 3 Pro}

\begin{tcolorbox}[colback=blue!3!white,
                  colframe=blue!75!black,
                  title=Example1,
                  fonttitle=\bfseries,
                  coltitle=white,
                  sharp corners=south,
                  breakable]

\begin{center}
\includegraphics[width=0.6\linewidth]{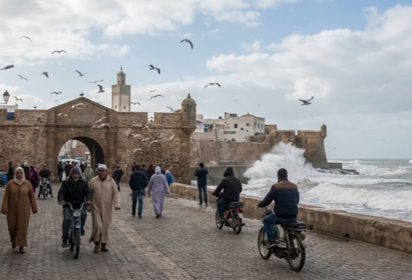} 
\end{center}

\{

“reason": "The audio recording provides a distinct sonic fingerprint pointing to the Maghreb region of North Africa. The most prominent initial clue is the Adhan (Islamic call to prayer) performed in a style typical of the region, immediately placing the location in a Muslim-majority country. This is overlaid with the loud, persistent squawking of seagulls and significant wind noise, which strongly suggests a coastal city on a major body of water. The traffic noise includes the distinctive high-pitched whine of 2-stroke mopeds or scooters navigating narrow streets, a common soundscape in Moroccan medinas. In the second half of the recording, the sound shifts to powerful, rhythmic waves crashing against a shore, sounding more like the heavy swells of the Atlantic Ocean than the typically calmer Mediterranean. Linguistically, shouts of 'Yallah' (Arabic) and what appears to be French phrases are audible, reflecting the bilingual heritage of Morocco. The specific combination of intense wind, overwhelming seagull presence, the Adhan, and heavy Atlantic surf is the hallmark of Essaouira, Morocco, famously known as the 'Wind City of Africa' with its historic fortified port."

"city": "Essaouira",    

"country": "Morocco",    

"continent": "Africa",    

"longitude": -9.7595,   

"latitude": 31.5085

\}

\end{tcolorbox}

\subsubsection{Example 1-GPT-4o Audio}

\begin{tcolorbox}[colback=blue!3!white,
                  colframe=blue!75!black,
                  title=Example 1,
                  fonttitle=\bfseries,
                  coltitle=white,
                  sharp corners=south,
                  breakable]

\begin{center}
\includegraphics[width=0.6\linewidth]{figures/appendix/case2.png} 
\end{center}

\{

“reason": "The audio recording features a clear Adhan, the Islamic call to prayer, which is typically heard in Muslim-majority countries. The background noises include traffic and urban sounds, suggesting a city environment. The presence of the Adhan combined with the urban noise leads to the conclusion that this might be a city in a country where Islam is predominant. Considering well-known cities, Istanbul in Turkey is a plausible guess due to its size, significance in the Muslim world, and blend of modernity and tradition. The city is often busy with traffic, as reflected in the audio, and the geographical location aligns with these audible cues."

"city": "Istanbul",    

"country": "Turkey",    

"continent": "Europe",    

"longitude": 28.9784,   

"latitude": 41.0082

\}

\end{tcolorbox}

\subsubsection{Example 1-Qwen3-Omni}

\begin{tcolorbox}[colback=blue!3!white,
                  colframe=blue!75!black,
                  title=Example 1,
                  fonttitle=\bfseries,
                  coltitle=white,
                  sharp corners=south,
                  breakable]

\begin{center}
\includegraphics[width=0.6\linewidth]{figures/appendix/case2.png} 
\end{center}

\{

“reason": "The audio contains a mix of sounds that strongly suggest a specific geographical location. The most prominent sound is a loud, high-pitched, and sustained horn, characteristic of a Japanese train or subway announcement. This is followed by a rhythmic, metallic clatter of wheels on tracks, typical of a train moving on rails. The background also features the sound of a large, powerful engine, possibly from a ferry or a large boat, with the sound of waves. The combination of these sounds—train/subway, ferry, and ocean waves—points to a coastal city with a well-developed public transportation system. The language spoken in the background is Japanese. The combination of these elements, particularly the specific sound of the train horn and the coastal environment, is highly characteristic of Japan. The city of Kobe, Japan, is a major port city with a significant train network and is located on the coast of the Seto Inland Sea, which would produce the sounds of both a train and a ferry."

"city": "Kobe",    

"country": "Japan",    

"continent": "Asia",    

"longitude": 135.198,   

"latitude": 34.693

\}

\end{tcolorbox}

\subsubsection{Example 1-Mimo-Audio}

\begin{tcolorbox}[colback=blue!3!white,
                  colframe=blue!75!black,
                  title=Example 1,
                  fonttitle=\bfseries,
                  coltitle=white,
                  sharp corners=south,
                  breakable]

\begin{center}
\includegraphics[width=0.6\linewidth]{figures/appendix/case2.png} 
\end{center}

\{

“reason": "The audio contains several distinct clues that point to a specific geographical location. The primary clue is the prominent sound of a train horn, which is repeated multiple times throughout the recording. This suggests the recording was made near a railway line. Additionally, there are background sounds of traffic, including cars and possibly buses, indicating an urban environment. The ambient noise level is high, with a mix of mechanical and human-made sounds, which is typical of a bustling city. While there are no clear spoken words or languages that could definitively identify the location, the combination of train sounds and urban traffic noise strongly suggests a major city with an active rail network. One possible candidate is Mumbai, India, which is known for its dense urban environment and extensive railway system. The sounds of trains and traffic are consistent with the city's infrastructure. However, without more specific audio cues, this remains an educated guess."

"city": "Mumbai",    

"country": "India",    

"continent": "Asia",    

"longitude": 72.8,   

"latitude": 19.0

\}

\end{tcolorbox}

\subsubsection{Example 2-Gemini 3 Pro}

\begin{tcolorbox}[colback=blue!3!white,
                  colframe=blue!75!black,
                  title=Example 2,
                  fonttitle=\bfseries,
                  coltitle=white,
                  sharp corners=south,
                  breakable]

\begin{center}
\includegraphics[width=0.6\linewidth]{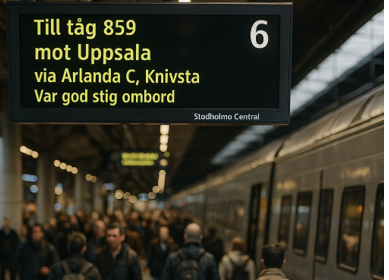} 
\end{center}

\{

"reason": "The audio environment is a large, reverberant hall with the distinct sounds of rolling suitcases, closing doors, and murmuring crowds, indicating a busy train station. At 0:48, faint conversation in Swedish can be heard. The definitive geographical evidence begins at 1:08 with a standard electronic chime followed by a female automated station announcement in Swedish. The voice announces: 'Till tåg [number]... mot Uppsala... via Arlanda C, Knivsta... Var god stig ombord, spår 6' (To train... towards Uppsala... via Arlanda Central, Knivsta... Please board, track 6). This specific route—traveling northbound to the city of Uppsala with stops at Arlanda Airport (Arlanda C) and Knivsta—originates from the main railway hub in the region. Therefore, this is recorded at Stockholm Central Station, specifically on the platform level.", 

"city": "Stockholm",   

"country": "Sweden",   

"continent": "Europe",  

"longitude": 18.0583,   

"latitude": 59.3301

\}

\end{tcolorbox}

\subsubsection{Example 2-Qwen3-Omni}

\begin{tcolorbox}[colback=blue!3!white,
                  colframe=blue!75!black,
                  title=Example 2,
                  fonttitle=\bfseries,
                  coltitle=white,
                  sharp corners=south,
                  breakable]

\begin{center}
\includegraphics[width=0.6\linewidth]{figures/appendix/case3.png} 
\end{center}

\{

"reason": "The audio contains several key clues pointing to a specific location. The most definitive clue is the announcement made in Korean, which is a language spoken in Korea. The announcement is for a 'KTX' train, which is the high-speed rail service operated by Korail in South Korea. The announcement also mentions 'Seoul Station', which is a major railway hub in Seoul. The background sounds include the distinct rumbling and clatter of a train moving on tracks, the sound of doors closing, and the general ambiance of a train station. The combination of the Korean language, the mention of KTX and Seoul Station, and the train sounds strongly indicates that this recording was made at Seoul Station in South Korea.", 

"city": "Seoul",   

"country": "South Korea",   

"continent": "Asia",  

"longitude": 126.9781,   

"latitude": 37.5326

\}

\end{tcolorbox}

\subsubsection{Example 3-Gemini 3 Pro}

\begin{tcolorbox}[colback=blue!3!white,
                  colframe=blue!75!black,
                  title=Example 3,
                  fonttitle=\bfseries,
                  coltitle=white,
                  sharp corners=south,
                  breakable]

\begin{center}
\includegraphics[width=0.6\linewidth]{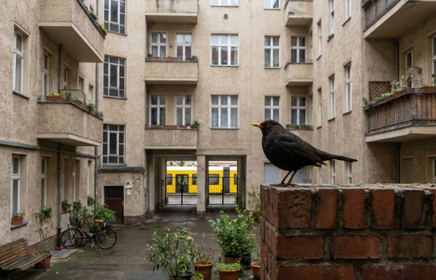} 
\end{center}

\{    

"reason": "The most prominent audio clue is the distinct, fluty, and melodious song of a Common Blackbird (Turdus merula). This bird is widespread across Europe and is famous for its adaptability to urban environments. The acoustic characteristics of the recording are equally telling: there is a noticeable reverberation or echo to the bird's song and the background noises, which strongly suggests the recording was taken in a 'Hinterhof'—a large, enclosed inner courtyard formed by block-style apartment buildings. This architectural style is iconic to Central European cities, particularly Berlin. In the background, there is a steady 'city hum' of distant traffic without aggressive honking, and the mechanical sounds at the beginning (a heavy rumble and metallic squeal) resemble a tram (Straßenbahn) or urban train passing nearby. The combination of the specific 'courtyard acoustics,' the pervasive blackbird song, and the sounds of European public transport infrastructure creates a sonic signature that is widely recognized as the soundscape of a residential neighborhood in Berlin, Germany.",

"city": "Berlin"

"country": "Germany",

"continent": "Europe",  

"longitude": 13.405, 

"latitude": 52.52

\}

\end{tcolorbox}

\subsubsection{Example 3-GPT-4o Audio}

\begin{tcolorbox}[colback=blue!3!white,
                  colframe=blue!75!black,
                  title=Example 3,
                  fonttitle=\bfseries,
                  coltitle=white,
                  sharp corners=south,
                  breakable]

\begin{center}
\includegraphics[width=0.6\linewidth]{figures/appendix/case4.png} 
\end{center}

\{    

"reason": "The audio recording contains ambient nature sounds, with prominent bird calls and chirping that seems specific to a suburban or park area. At points, there are also distant traffic noises, indicating a nearby roadway or urban environment. The type of birds and the overall soundscape suggest a temperate climate, possibly in the northern hemisphere. Given the generic nature sounds without distinctive markers like specific language or cultural audio indicators, a specific location prediction is difficult. However, based on the temperate nature, and commonality of such parks and environments, I am considering a well-known global city with expansive parks and temperate climate conditions. This leads me to a plausible guess of New York City, with Central Park being an exemplary location for such an environment. This choice is speculative and based on the assumption of a large city with notable green spaces and a presence of city and nature sounds.",

"city": "New York City"

"country": "United States",

"continent": "North America",  

"longitude": -73.9654, 

"latitude": 40.7829

\}

\end{tcolorbox}

\subsubsection{Example 3-Qwen3-Omni}

\begin{tcolorbox}[colback=blue!3!white,
                  colframe=blue!75!black,
                  title=Example 3,
                  fonttitle=\bfseries,
                  coltitle=white,
                  sharp corners=south,
                  breakable]

\begin{center}
\includegraphics[width=0.6\linewidth]{figures/appendix/case4.png} 
\end{center}

\{    

"reason": "The audio contains a prominent and complex bird song, characteristic of a song thrush (Turdus philomelos), which is common in Europe. The background is dominated by the sounds of urban traffic, including cars and a distinct 'honk' of a vehicle horn. The combination of these specific bird calls and the nature of the traffic noise strongly suggests a location in Western Europe. The overall soundscape is typical of a city park or a green space within a European city. The specific combination of the song thrush's song and the urban environment points towards a location in the United Kingdom, where this bird is widespread and the traffic patterns are consistent with the audio. London is a major city in the UK with such environments.",

"city": "London"

"country": "United Kingdom",

"continent": "Europe",  

"longitude": -0.1278, 

"latitude": 51.5074

\}

\end{tcolorbox}

% TODO
% Instructions Given To Participants
% Recruitment And Payment
% Data Consent
% Characteristics Of Annotators

\end{document}